\documentclass[prb,onecolumn,showpacs,amssymb,superscriptaddress]{revtex4-2}
\usepackage{graphicx}
\usepackage{amsmath}
\usepackage{amsfonts}
\usepackage{amsthm}
\usepackage{amssymb}
\usepackage{amsbsy}
\usepackage{wasysym}
\usepackage{bm}
\usepackage{bbm}
\usepackage{mathrsfs}
\usepackage{color}
\usepackage{hyperref}
\usepackage{braket}
\usepackage{soul}
\usepackage{mathtools}
\usepackage{soul}
\usepackage[symbol]{footmisc}

\setcitestyle{super}

\newcommand*{\citen}[1]{%
  \begingroup
    \romannumeral-`\x 
    \setcitestyle{numbers}%
    \cite{#1}%
  \endgroup   
}

\date{\today}
\begin{document}
\title{Hydrodynamics of spin transport in the Haldane-Shastry chain}

\author{Vir B. Bulchandani}

\affiliation{Department of
Physics and Astronomy, Rice University, 6100 Main Street
Houston, TX 77005, USA}\thanks{on leave}

\affiliation{Institut f{\"u}r Theoretische Physik, Leibniz Universit{\"a}t Hannover, Appelstra{\ss}e 2, 30167 Hannover, Germany}

\author{Hyunsoo Ha}

\affiliation{Department of Physics, Princeton University, Princeton, New Jersey 08544, USA}

\begin{abstract}
We use kinetic theory and the thermodynamic Bethe ansatz to derive the hydrodynamics of spin transport in the Haldane-Shastry chain. We obtain analytical expressions for relaxation from weak-domain-wall initial conditions and corresponding predictions for the spin Drude weight. We corroborate the latter against an exact microscopic calculation at infinite temperature and Luttinger-liquid expectations at zero temperature. We further obtain an analytical expression for the long-wavelength dynamics of the longitudinal spin structure factor at all fillings and temperatures. We show that the latter interpolates between an empirically exact expression that we have found at infinite temperature and the long-wavelength limit of the Haldane-Zirnbauer formula at zero temperature.
\end{abstract}
\maketitle

\tableofcontents
\section{Introduction} 
The Haldane-Shastry spin chain famously supports quasiparticles with a free-particle-like dispersion~\cite{HaldaneOrig,ShastryOrig}. The interpretation of the Haldane-Shastry chain as a non-interacting system is further supported by several results~\cite{haldane1991spinon,EsslerSpinon,NoSpinonAttraction} relating the spectrum and dynamics of the Haldane-Shastry chain to the properties of a non-interacting ``spinon gas''. However, a microscopic mapping from the Haldane-Shastry Hamiltonian to a spinon gas Hamiltonian remains elusive~\cite{TalstraSpinon}. The analogy with the spinon gas is nevertheless sufficiently precise that it gives rise to exact closed-form expressions for various thermodynamic properties of the Haldane-Shastry chain~\cite{haldane1991spinon,HaHaldane,haldane1994physics}, as well as its zero-temperature dynamical spin structure factor~\cite{HaldaneZirnbauer,HaldaneTalstra}.

The dynamics of this model bears revisiting in light of subsequent advances in the far-from-equilibrium dynamics of quantum integrable systems. The kinetic theory of short-range interacting integrable systems is now reasonably well understood~\cite{Castro_Alvaredo_2016,Bertini_2016} (see e.g. the collection of review articles Ref.~\citen{Bastianello_2022} and a recent book~\cite{spohn2024hydrodynamic}) with well-established antecedents in the theory of classical soliton gases~\cite{percus1969exact,zakharov1971kinetic,el2005kinetic}. The resulting theory is usually referred to as ``generalized hydrodynamics'' in recognition of the infinite set of hydrodynamic equations that it implies~\cite{Castro_Alvaredo_2016,Bertini_2016}, while its underlying transport equation more closely resembles a kinetic theory of quasiparticles~\cite{Bulchandani_2018,Doyon_2018}. This formalism was recently applied to the long-range interacting quantum and classical Calogero models~\cite{Bulchandani_2021}. In the latter case, a kinetic equation can be derived directly from the dynamics of the Lax matrix and exhibits excellent agreement with numerical simulations of the microscopic dynamics. Moreover, the quasiparticle kinetic theories for both the quantum and classical Calogero models take the form of free-streaming Boltzmann equations, which is consistent with the interpretation of these models as gases of free anyons\cite{PolychronakosReview}.

It is tempting to expect that the Haldane-Shastry chain should be amenable to the same approach. However, the regime of validity of such kinetic theories is less clear for systems with global continuous symmetries, such as the global $SU(2)$ symmetry that characterizes both the Haldane-Shastry chain and the better-studied Heisenberg chain. For the Heisenberg chain, the possibility arises of anomalous spin transport that is neither ballistic nor diffusive~\cite{ZnidOrig,KPZ,bulchandani2021superdiffusion,scheie2021detection,wei2022quantum,rosenberg2024dynamics}, and has so far only been treated theoretically with self-consistent (albeit empirically accurate~\cite{KPZtest}) scaling arguments~\cite{GopVass,Ilievski_2021}. Moreover, for any integrable model with a global continuous symmetry, a conventional hydrodynamic theory should be able to capture time evolution from locally equilibrated initial conditions involving Goldstone modes of that symmetry, but describing such time evolution in detail seems to lie beyond all existing theoretical approaches~\cite{SoftGauge,bulchandani2021superdiffusion}. 

For the Haldane-Shastry chain, we can nevertheless hope to partially circumvent these difficulties because spin transport in this model is purely ballistic in the thermodynamic limit~\cite{Sirker_2011}. This is ultimately a consequence~\cite{Bernevig01} of the model's Yangian symmetry, which yields a conserved charge that is proportional to the spin current in the thermodynamic limit~\cite{Yangian,haldane1994physics} and therefore rules out the possibility of anomalous spin transport with its attendant theoretical difficulties. While all available theoretical evidence~\cite{Bastianello_2022,spohn2024hydrodynamic} suggests that ballistic transport in integrable systems is captured exactly by the techniques of generalized hydrodynamics, this paper represents the first application of such techniques to a long-range interacting spin chain. We will therefore verify our results against independent microscopic calculations wherever possible, obtaining several new results on spin transport in the Haldane-Shastry chain in the process.

The paper is structured as follows. We first summarize the derivation~\cite{Bernevig01,Sirker_2011} of purely ballistic spin transport for the Haldane-Shastry chain and review some general strategies for obtaining dynamical structure factors\cite{kubo2012statistical} and Drude weights~\cite{Vasseur_2015,Ilievski_2017,Bulchandani_2018,DoyonSpohn} of local conserved charges from hydrodynamics. We then discuss the thermodynamics of the Haldane-Shastry chain and derive several apparently new expressions characterizing the equilibrium root densities for this model (Eqs. \eqref{eq:totaldos}, \eqref{eq:exactdos}, \eqref{eq:weightedsumrule} and \eqref{eq:sumrule}) that are required to reconcile the thermodynamic Bethe ansatz (TBA) treatment of its thermodynamics with earlier predictions from the spinon-gas picture~\cite{haldane1991spinon,haldane1994physics}. We use this thermodynamic Bethe ansatz to derive a kinetic theory of quasiparticles in the Haldane-Shastry model following standard techniques~\cite{Castro_Alvaredo_2016,Bertini_2016} and show that at the level of the magnetization, this reduces to a much simpler kinetic theory of a ``signed spinon density'', which in turn yields predictions for the physical spin density and spin current density at each spacetime point.

This simplification allows us to derive explicit formulae for the hydrodynamics of spin in the Haldane-Shastry model, which we apply to studying time evolution from weak domain walls, to the evaluation of spin Drude weights, and to the long-wavelength dynamics of the dynamical structure factor. While our results apply in principle to all temperatures and filling fractions, we focus on two analytically tractable cases in particular, namely half filling at arbitrary temperature and infinite temperature at arbitrary filling. Where possible, we check our analytical predictions against microscopically exact results and find agreement in all cases. We conclude with a discussion of some theoretical questions raised by our analysis.

\section{Background}
We will focus on the Haldane-Shastry Hamiltonian~\cite{HaldaneOrig,ShastryOrig} in the form
\begin{equation}
\label{eq:HS}
\hat{H} = J\sum_{\substack{x,y=1,\\x<y}}^L \frac{\hat{\mathbf{S}}_x \cdot \hat{\mathbf{S}}_y}{d(x,y)^2} - 2h \sum_{x=1}^L \hat{S}^z_x
\end{equation}
on a ring of $L$ sites in an external $z$-field, where the chord distance between any given pair of spins $d(x,y) = \frac{L}{\pi} |\sin{\pi(x-y)/L}|$, and $\hat{\mathbf{S}}_n = (\hat{S}_n^x,\hat{S}_n^y,\hat{S}_n^z)$ denote the usual $S=1/2$ operators at lattice site $n$, with $[\hat{S}_m^\alpha,\hat{S}_n^\beta] = i \delta_{mn}\epsilon^{\alpha \beta \gamma} \hat{S}_n^\gamma$.
\subsection{Ballistic spin transport from Yangian symmetry}
Spin transport in the Haldane-Shastry model is known~\cite{Sirker_2011} to be ballistic in the thermodynamic limit $L\to \infty$, because the spin current coincides~\cite{Bernevig01} with a conserved generator of the so-called ``Yangian'' symmetry~\cite{Yangian,haldane1994physics} of the model. The external magnetic field in Eq. \eqref{eq:HS} breaks the full $SU(2)$ and Yangian symmetries of $\hat{H}$ whenever it is non-zero, and the surviving Yangian generator or ``rapidity operator'' can be written as
\begin{equation}
\hat{\Lambda} = \frac{\pi}{L} \sum_{x \neq y} \cot{\left(\frac{\pi (x-y)}{L}\right)} \, \mathbf{z} \cdot  \hat{\mathbf{S}}_x \times \hat{\mathbf{S}}_y
\end{equation}
where $\mathbf{z} = (0,0,1)$ denotes the $z$ basis vector in spin space. The rapidity operator is exactly conserved, $[\hat{H},\hat{\Lambda}]=0$. We now briefly relate this to the spin current. To this end, let us pass to the thermodynamic limit with $L \to \infty$ and use the shorthand $J_{|x-y|} = \frac{J}{|x-y|^2}$ with $J_0=0$. We first note that the Heisenberg equations of motion for $\hat{S}^z_x$ can be written as the discrete continuity equation
\begin{equation}
\partial_t \hat{S}^z_x + \hat{j}^S_{x+1}-\hat{j}^S_{x} = 0,
\end{equation}
where the local spin current operators
\begin{equation}
j^S_x = \sum_{r>0} \sum_{y=x-r}^{x-1} J_r \mathbf{z} \cdot  \hat{\mathbf{S}}_y \times \hat{\mathbf{S}}_{y+r}.
\end{equation}
It follows that the total spin current operator $\hat{J}^S = \sum_{x=-\infty}^\infty \hat{j}^S_x$ is given by
\begin{equation}
\hat{J}^S = -\frac{1}{2} \sum_{x,y=-\infty}^{\infty} (x-y) J_{|x-y|}\mathbf{z} \cdot \hat{\mathbf{S}}_x \times \hat{\mathbf{S}}_y = -\frac{J}{2} \sum_{x,y=-\infty}^\infty \frac{1}{x-y}\mathbf{z} \cdot \hat{\mathbf{S}}_x \times \hat{\mathbf{S}}_y = -\frac{J}{2}\hat{\Lambda},
\end{equation}
and therefore that $[\hat{H},\hat{J}^S]=0$ in the thermodynamic limit. 

\subsection{Dynamical structure factors from hydrodynamics}
\label{sec:DSF}
We now summarize the formal connection between dynamical structure factors and hydrodynamics, which follows from the results of Kubo~\cite{Kubo1957,kubo2012statistical,Bertini2021RMP}. Thus let $\hat{H}$ be a local translation-invariant lattice Hamiltonian on a ring of $L$ sites, $\hat{O}_x$ be the density of a local conserved charge $\hat{O} = \sum_{x=1}^L \hat{O}_x$ and $\hat{\rho}_0=\exp(-\beta\hat{H})/Z$ the equilibrium density matrix at inverse temperature $\beta$. Consider a weak modulation of this equilibrium density matrix, of the form
\begin{equation}
\label{eq:perturbedDM}
    \hat{\rho}_\epsilon = \frac{1}{Z(\epsilon)}\exp\left(-\beta\left(\hat{H}-\epsilon\sum_{y=1}^L f(y)\hat{O}_{y} \right)\right),
\end{equation}
where $Z(\epsilon) = \mathrm{tr}\left[\exp\left(-\beta\left(\hat{H}-\epsilon\sum_{y=1}^L f(y)\hat{O}_{y} \right)\right)\right]$. For a given operator $\hat{A}$ we will write $\langle \hat{A} \rangle_{\epsilon} = \mathrm{tr}[\hat{\rho}_{\epsilon} \hat{A}]$ and use the notation $\hat{A}(t) = e^{i\hat{H}t} \hat{A}e^{-i\hat{H}t}$ for Heisenberg picture dynamics. It will also be useful to write $\Delta \hat{A} =  \hat{A}  - \langle \hat{A} \rangle_0$. Our goal is to compute the dynamical structure factor 
\begin{equation}
    C(x-y,t) = \langle \Delta \hat{O}_x(t) \Delta \hat{O}_y\rangle_0.
\end{equation} 
The motivation for considering the initial condition Eq. \eqref{eq:perturbedDM} is that for sufficiently slowly varying $f(x)$, $\hat{\rho}_{\epsilon}$ is locally in thermal equilibrium and the time evolution of $\langle \hat{O}_x(t) \rangle_{\epsilon}$ should be obtainable from hydrodynamics. At the same time, the time evolution of $\langle \hat{O}_x(t) \rangle_{\epsilon}$ is determined by $C(x-y,t)$ at leading order in $\epsilon$. Comparing these expressions yields a hydrodynamic prediction for the dynamical structure factor~\cite{Bertini2021RMP}.

In more detail, expanding $\hat{\rho}_\epsilon$ to linear order in $\epsilon$ yields\cite{Kubo1957}
\begin{equation}
\label{eq:initstate}
    \hat{\rho}_\epsilon = \hat{\rho}_0\left[\hat{\mathbbm{1}} + \epsilon \sum^L_{y=1} f(y) \int_0^\beta d\lambda \, e^{\lambda \hat{H}} \Delta \hat{O}_{y} e^{-\lambda \hat{H}} + 
    \mathcal{O}(\epsilon^2)\right].
\end{equation}
It follows that
\begin{align}
\langle \Delta \hat{O}_x(t) \rangle_{\epsilon} 
\label{eq:KCFfromhydro}
= \beta \epsilon \sum_{y=1}^L  K(x-y,t)f(y) + \mathcal{O}(\epsilon^2)
\end{align}
where $K(x-y,t)$ denotes the Kubo correlation function $K(x-y,t) \coloneqq \frac{1}{\beta} \int^\beta_0 d\lambda \, \langle \Delta \hat{O}_x(t+i\lambda) \Delta \hat{O}_y\rangle_{0}$. 

For a controlled comparison with hydrodynamics, we denote Fourier transforms in space and time by the arguments $q$ and $\omega$ respectively, i.e. use the notation $F(q,t) = \sum_{y=1}^L e^{-iq(x-y)}F(x-y,t)$ and $F(x,\omega) = \int_{-\infty}^{\infty} dt \, e^{i\omega t} F(x,t)$, and make the specific choice of initial condition $f(x) = e^{iqx}$ (which should be viewed as a formal trick for computing the structure factor rather than a physical initial condition). Then
\begin{equation}
\label{eq:KuboCF}
K(q,t) = \frac{1}{\beta \epsilon} e^{-iqx}\langle \Delta \hat{O}_x(t) \rangle_{\epsilon}
\end{equation}
and the dynamical structure factor can be obtained from the Kubo correlation function using the formula\cite{Kubo1957}
\begin{equation}
\label{eq:FourierRel}
K(q,\omega) = \left(\frac{1-e^{-\beta \omega}}{\beta \omega}\right) C(q,\omega).
\end{equation}
Hydrodynamics is expected to yield accurate predictions for the right-hand side of Eq. \eqref{eq:KuboCF} in the limit $q\ll 1$. Eq. \eqref{eq:FourierRel} then yields a prediction for the long-wavelength behaviour of the dynamical structure factor. 

\subsection{The spin Drude weight of the Haldane-Shastry chain}
\label{sec:SDW}
Because the total spin current of the Haldane-Shastry chain is conserved in the thermodynamic limit, the real part of the AC spin conductivity
\begin{equation}
\label{eq:defDrude}
\sigma'(\omega) = \pi \beta D^S \delta(\omega)
\end{equation}
consists solely of a Drude peak in the thermodynamic limit, with Drude weight~\cite{kubo2012statistical,Zotos_1997} 
\begin{equation}
\label{eq:microDrude}
D^S =  \lim_{L \to \infty} \frac{\langle (\hat{J}^S)^2 \rangle_0 }{L} =  \lim_{L \to \infty} \frac{J^2\langle \hat{\Lambda}^2 \rangle_0}{4L}
\end{equation}
in the notation of previous sections. Such equilibrium expectation values of operators are difficult to compute analytically in general, though an exact result can be obtained at infinite temperature and is given in Eq. \eqref{eq:infTexact}. Note that our conventions in Eq. \eqref{eq:defDrude} ensure a non-vanishing Drude weight at infinite temperature, but yield a vanishing Drude weight at zero temperature. At low temperatures, we will instead consider the rescaled quantity $\tilde{D}^S = \beta D^S$.

An alternative, hydrodynamic method for computing spin Drude weights consists of studying time evolution from the ``weak domain wall'' initial condition, that has proved instrumental for understanding spin transport in the Heisenberg chain~\cite{Vasseur_2015,Ilievski_2017,Bulchandani_2018,DoyonSpohn,KPZ}. Thus we consider two semi-infinite reservoirs at the same inverse temperature $\beta$, joined at $x=0$ and prepared in equilibrium with magnetic fields $h_L = h+\epsilon$ and $h_R = h-\epsilon$ for the left and right reservoirs respectively, write $\tilde{\epsilon} = \beta \epsilon$ and assume $|\tilde{\epsilon}| \ll 1$. The resulting initial magnetic field profile can be written as
\begin{equation}
\label{eq:twores}
h(x) = h+\epsilon(1-2\Theta(x)),
\end{equation}
where $\Theta(x)$ denotes the Heaviside step function. Letting $\hat{\rho}_{\epsilon}$ denote the corresponding density matrix, as in the previous section, a hydrodynamic expression for the spin Drude weight is given by~\cite{Vasseur_2015,Ilievski_2017,Bulchandani_2018,DoyonSpohn}
\begin{equation}
\label{eq:hydroDrude}
D^S = \lim_{\epsilon \to 0} \lim_{t\to\infty} \frac{1}{4\tilde{\epsilon} t} \sum_{x=-\infty}^{\infty} \langle \hat{j}^S_x(t) \rangle_{\epsilon}.
\end{equation}
Our strategy in this paper will be to use hydrodynamics to compute $D^S$ analytically from Eq. \eqref{eq:hydroDrude}, and then to check the latter by direct evaluation of the microscopic formula Eq. \eqref{eq:microDrude} where possible. 

\section{Thermodynamics of the Haldane-Shastry model revisited}
We now revisit the thermodynamics of the Haldane-Shastry model. This was first derived in the language of the ``spinon gas'' in Ref.~\citen{haldane1991spinon}, and subsequently given a more conventional thermodynamic Bethe ansatz treatment in Ref.~\citen{HaHaldane}. However, to the best of our knowledge, the thermodynamic Bethe ansatz and spinon gas treatments have never been reconciled with one another in print. In this section, we will follow the thermodynamic Bethe ansatz route, in the process deriving several apparently new formulae that are needed to establish consistency with the earlier spinon gas results.

\subsection{Thermodynamic Bethe ansatz equations}
Recall that the quasiparticle content of the Haldane-Shastry model can be expressed in terms of infinitely many ``string'' excitations in the thermodynamic limit\cite{HaHaldane}, which consist of superpositions of $n \in \mathbb{N}$ flipped spins relative to a ferromagnetic ``pseudovacuum'' $|\Omega\rangle = |\uparrow \uparrow \ldots \uparrow \rangle$. Let $\rho_{n}(k)$ denote the occupied densities of states for these strings, with $n=1,2,\ldots$ the string index counting the number of flipped spins and $k \in [-\pi,\pi)$ the pseudomomentum of each string (for convenience this is $\pi$ less than the crystal momentum\cite{HaHaldane}). We also let $\rho^t_n(k)$ denote the total density of states and $\rho^h_n(k)$ the density of holes, so that $\rho^t_n(k) = \rho_n(k) + \rho^h_n(k)$. At various points we will use the notations $\eta_n(k) = \rho^h_n(k)/\rho_n(k)$ and $\theta_n(k) = \rho_n(k)/\rho^t_n(k) = (1+\eta_n(k))^{-1}$, and largely suppress the explicit $k$ dependence. Then the bare dispersion of the $n$th string relative to the pseudovacuum is given by
\begin{equation}
e_n(k) = e_0(k) + 2nh, \quad n \in \mathbb{N},
\end{equation}
where
\begin{equation}
e_0(k) = \frac{J}{4}(k^2-\pi^2)
\end{equation}
and the Bethe equations take the form\cite{takahashi2005thermodynamics,HaHaldane}
\begin{equation}
\label{eq:BetheEq}
\rho^t_{n}(k) + \sum_{m=1}^\infty C_{n,m} \rho_{m}(k) = \frac{1}{2\pi}, \quad n \in \mathbb{N},
\end{equation}
where
\begin{equation}
C_{n,m} = 2 \min(n,m) - \delta_{n,m}.
\end{equation}
For states in thermal equilibrium at temperature $T$, the $\eta_n$ satisfy the Yang-Yang equations
\begin{equation}
\label{eq:YangYang}
\log{\eta_n}(k) = \beta e_n(k) + \sum_{m=1}^\infty C_{n,m} \log{(1+\eta_m^{-1}(k))}, \quad n \in \mathbb{N}.
\end{equation}
These can be written as\cite{HaHaldane,takahashi2005thermodynamics}
\begin{equation}
\label{eq:diffeq}
\eta_n^2 = (1+\eta_{n-1})(1+\eta_{n+1}), \quad n \in \mathbb{N},
\end{equation}
with boundary condition
\begin{equation}
\eta_0(k) = e^{\beta e_0(k)}-1
\end{equation}
for each $k$. The results above were previously obtained in Ref. \citen{HaHaldane}. If, following Haldane\cite{haldane1994physics}, we introduce a function $\mu(k)$ such that
\begin{equation}
\label{eq:defmu}
e^{\beta e_0(k)/2}\sinh{\beta h} = \sinh{\beta h \mu(k)},
\end{equation}
we can write the solution\cite{takahashi2005thermodynamics} to Eq. \eqref{eq:diffeq} as
\begin{equation}
\label{eq:valeta}
\eta_n(k) = \left( \frac{\sinh{\beta h(n+\mu(k))}}{\sinh{\beta h}}\right)^2 - 1.
\end{equation}
For future reference, we also note the following analytically tractable limits of Eq. \eqref{eq:defmu}:
\begin{equation}
\label{eq:tractlim}
\mu(k) = \begin{cases} 1, & \beta \to 0, \tilde{h} = \beta h = \mathrm{const}, \\
e^{\beta e_0(k)/2}, & h=0.
\end{cases}
\end{equation}

We now derive some apparently new expressions for the string densities at all fillings and temperatures that are needed to bridge the gap between the Bethe ansatz and the ``spinon gas'' descriptions of the Haldane-Shastry chain (Eqs. \eqref{eq:totaldos}, \eqref{eq:exactdos}, \eqref{eq:weightedsumrule} and \eqref{eq:sumrule}). First, following Takahashi, we differentiate the Yang-Yang equations Eq. \eqref{eq:YangYang} with respect to $J$ and find that in thermal equilibrium, the total density of states
\begin{equation}
\label{eq:totaldosfromTBA}
\rho^t_n(k) = \frac{1}{2\pi} \frac{J}{\beta e_0(k)} \frac{\partial_J \eta_n(k)}{\eta_n(k)}
\end{equation}
satisfies the Bethe equations Eq. \eqref{eq:BetheEq}. This yields the explicit expressions
\begin{equation}
\label{eq:totaldos}
\rho^t_n = \frac{1}{2\pi} \frac{\cosh{\beta h(n+\mu)}}{\cosh{\beta h \mu }}\frac{\sinh{\beta h \mu}\sinh{\beta h (n+\mu)}}{\sinh{\beta h(n+\mu-1) }\sinh{\beta h(n+\mu+1)}}
\end{equation}
and
\begin{equation}
\label{eq:exactdos}
\rho_n = \frac{1}{2\pi} \frac{\cosh{\beta h(n+\mu)}}{\cosh{\beta h \mu }}\frac{\sinh^2{\beta h}\sinh{\beta h \mu}}{\sinh{\beta h(n+\mu-1)}\sinh{\beta h (n+\mu)} \sinh{\beta h(n+\mu+1)}}.
\end{equation}

In terms of the root densities $\rho_n(k)$, the equilibrium magnetization density $s^z(\beta,h)$ can be written as
\begin{equation}
\label{eq:defmag}
s^z(\beta,h) = \lim_{L \to \infty} \frac{1}{L} \sum_{x = 1}^{L} \langle \hat{S}^z_x \rangle = \frac{1}{2} - \sum_{n=1}^\infty n \int_{-\pi}^{\pi} dk \, \rho_{n}(k)
\end{equation}
while the free-energy density is given by~\cite{HaHaldane,takahashi2005thermodynamics}
\begin{equation}
\label{eq:freeenergydens}
f(\beta,h) = \lim_{L \to \infty} - \frac{1}{L\beta} \log{\mathrm{tr}\left[e^{-\beta \hat{H}}\right]} = \frac{J\pi^2}{24} - h - \frac{1}{\beta} \sum_{n=1}^\infty \int_{-\pi}^\pi  \frac{dk}{2\pi} \, \log{(1+\eta_n^{-1}(k))},
\end{equation}
whose temperature-independent part accounts for the energy density of the pseudovacuum.

For later reference, it will be helpful to explicitly define the infinite matrix
\begin{equation}
M_{n,m}(k) = \delta_{n,m} + C_{n,m} \theta_m(k)
\end{equation}
and its inverse (the ``dressing operator'')
\begin{equation}
\label{eq:defdressing}
U_{n,m}(k) = (M^{-1}(k))_{n,m}.
\end{equation}
\subsection{From string to spinon thermodynamics}
For an alternative perspective on the magnetization density, it is useful to define the ``signed spinon density at pseudomomentum $k$'', given by
\begin{equation}
\varsigma(k) = \frac{1}{2\pi} - \sum_{n=1}^\infty 2 n \rho_{n}(k),
\end{equation}
in terms of which the spin density
\begin{equation}
s^z = \frac{1}{2} \int_{-\pi}^{\pi} dk\, \varsigma(k)    
\end{equation}
and the spin current density
\begin{equation}
j^S = \frac{1}{2} \int_{-\pi}^{\pi} dk\, (Jk/2)\varsigma(k)
\end{equation}
respectively. In more detail, $\varsigma(k)$ can be understood in the spinon language~\cite{haldane1991spinon} as the difference in occupancy between spin $+1/2$ spinon orbitals and spin $-1/2$ spinon orbitals at pseudomomentum $k$. A remarkable feature of the thermodynamic Bethe ansatz for the Haldane-Shastry chain is that the sum $\sum_{n=1}^\infty n \rho_n(k)$ simplifies drastically, since upon combining Eqs. \eqref{eq:YangYang} and \eqref{eq:totaldos}, it follows that
\begin{equation}
\label{eq:weightedsumrule}
\sum_{n=1}^\infty n \rho_n(k) = - \frac{1}{4\pi} \frac{J}{\beta e_0(k)} \lim_{n \to \infty} \partial_J \log{\left(e^{-\beta e_n(k)} (1+\eta_n(k))\right)}.
\end{equation}
Computing the right-hand side explicitly using Eq. \eqref{eq:valeta} then yields the expression
\begin{equation}
\label{eq:magicformula}
\varsigma(k) = \frac{1}{2 \pi} \tanh{\beta h \mu(k)}
\end{equation}
for the signed spinon density, which in turn implies the closed form
\begin{equation}
\label{eq:TBAmag}
s^z(\beta,h) = \int_{-\pi}^\pi \frac{dk}{4\pi} \tanh{\beta h \mu(k)}
\end{equation}
for the equilibrium magnetization (note that the spin current density $j^S(\beta,h) = 0$ in thermal equilibrium). It follows that the spin susceptibility is given by
\begin{equation}
\label{eq:magsusc}
\chi(\beta,h) = \frac{\partial}{\partial h} s^z(\beta,h) = \int_{-\pi}^\pi \frac{dk}{4 \pi }\frac{\beta \tanh{\beta h \mu(k)}}{\tanh{\beta h}\cosh^2{\beta h \mu(k)}}.
\end{equation}
At half filling, this recovers an expression that was obtained previously from the spinon-gas description~\cite{haldane1991spinon}.

We can similarly derive a simplified expression for the free-energy density Eq. \eqref{eq:freeenergydens}. To this end, note that the Yang-Yang equations Eq. \eqref{eq:YangYang} imply that~\cite{takahashi2005thermodynamics}
\begin{equation}
\label{eq:sumrule}
\sum_{n=1}^{\infty} \log{(1+\eta_n^{-1})} = \frac{1}{2}\left(\log{(1+\eta_1)}-\beta e_1\right),
\end{equation}
yielding the formula
\begin{equation}
f(\beta,h) = - \frac{J\pi^2}{24} - \frac{1}{\beta} \int_{-\pi}^\pi \frac{dk}{2\pi} \log{\left(\frac{\sinh{\beta h (1+\mu(k))}}{\sinh{\beta h}}\right)}
\end{equation}
whose temperature-dependent part was previously derived from the spinon-gas description~\cite{haldane1994physics}. A useful corollary of Eq. \eqref{eq:sumrule} for computing thermal expectation values of densities of conserved charges is
\begin{equation}
\label{eq:energysumrule}
\sum_{n=1}^\infty \rho_n(k)  = \frac{1}{4\pi} \frac{\sinh{\beta h}}{\cosh{\beta h \mu(k)} \sinh{\beta h (1+\mu(k))}}.
\end{equation}

The possibility of obtaining these closed-form expressions from the thermodynamic Bethe ansatz seems to directly reflect the existence of a simple underlying description of the Haldane-Shastry model in terms of non-interacting spinons. Analogous simplifications in the thermodynamic Bethe ansatz for other $SU(2)$-symmetric spin chains, such as the Heisenberg chain, only occur at infinite temperature~\cite{takahashi2005thermodynamics}.

\subsection{Kinetic theory of the Haldane-Shastry model}
In the kinetic theory of integrable systems, the occupied densities of states $\rho_n(k)$ are promoted to space-time dynamical variables $\rho_n(x,t,k)$. At the ballistic scale, this yields the kinetic equations\cite{Bertini_2016,Castro_Alvaredo_2016}
\begin{equation}
\partial_t \rho_{n} + \partial_x (v^{\mathrm{eff}}_{n}[\rho] \rho_{n}) = 0, \quad n \in \mathbb{N}.
\end{equation}
For the Haldane-Shastry model, the group velocity for each quasiparticle species
\begin{equation}
\label{eq:veff}
v_n^{\mathrm{eff}}[\rho](k) = \frac{\sum_{m=1}^\infty U_{n,m}(k) e_m'(k)}{\sum_{m=1}^\infty U_{n,m}(k)} = e'_n(k) = \frac{Jk}{2}, \quad n \in \mathbb{N}.
\end{equation}
Thus all the quasiparticles in the model travel at their bare velocity. This is a consequence of the strict $k$-space locality of the quasiparticle dressing, which is certainly not the case in general~\cite{Castro_Alvaredo_2016,Bertini_2016}. 

We deduce that unlike for the Heisenberg chain~\cite{Bertini_2016}, all strings travel with the same group velocity $v(k) =  Jk/2$, and the kinetic theory assumes the non-interacting form
\begin{equation}
\label{eq:BB}
\partial_t \rho_n + (Jk/2) \partial_x \rho_n = 0, \quad n\in \mathbb{N}.
\end{equation}
This implies a corresponding kinetic equation for the signed spinon density, now promoted to a dynamical variable $\varsigma(x,t,k)$, namely
\begin{equation}
\label{eq:spinonbb}
\partial_t \varsigma + (Jk/2) \partial_x \varsigma = 0.
\end{equation}

At this point two observations are in order. First, the absence of velocity dressing in these equations also characterizes the closely related quantum Calogero model\cite{PolychronakosReview}, whose quasiparticle kinetic theory is the free-streaming Boltzmann equation\cite{Bulchandani_2021}. Second, the simple form of the dispersion, combined with the absence of velocity dressing, suggests that these equations are not susceptible to the dissipative\cite{DeNardis2018,Gopalakrishnan2018} corrections that are generic for other integrable models, so that the Haldane-Shastry model is non-interacting in the sense of Spohn\cite{SpohnIntNonInt}. The non-interacting form of Eq. \eqref{eq:spinonbb} is thus consistent with previous interpretations of the Haldane-Shastry model as a non-interacting spinon gas\cite{haldane1991spinon,EsslerSpinon,NoSpinonAttraction}. 

The equations Eq. \eqref{eq:BB} and \eqref{eq:spinonbb} are immediately solvable by characteristics, and the initial value problem at $t=0$ has the solution
\begin{equation}
\rho_n(x,t,k) = \rho_n(x-Jkt/2,0,k),
\end{equation}
which implies
\begin{equation}
\label{eq:stimeevo}
\varsigma(x,t,k) = \varsigma(x-Jkt/2,0,k).
\end{equation}
In particular, the ballistic-scale dynamics of the local magnetization density $s^z(x,t)$ is given by
\begin{equation}
\label{eq:magtimeevo}
s^z(x,t) = \frac{1}{2} \int_{-\pi}^\pi dk \, \varsigma(x-Jkt/2,0,k)
\end{equation}
while the spin current density
\begin{equation}
j^S(x,t) = \frac{1}{2} \int_{-\pi}^\pi dk \, (Jk/2)\varsigma(x-Jkt/2,0,k).
\end{equation}
In what follows, it will be useful to write $v_s = J \pi/2$ for the speed of the fastest quasiparticles in the system. By Eq. \eqref{eq:BB}, $v_s$ sets the speed limit for ballistic transport in the Haldane-Shastry chain at \textit{all} fillings and temperatures, not just at zero temperature where it coincides with the spinon velocity\cite{HaldaneOrig,ShastryOrig}.

\section{Hydrodynamics of spin transport in the Haldane-Shastry chain}
We now use Eqs. \eqref{eq:magtimeevo} and Eq. \eqref{eq:magicformula} to study various hydrodynamic regimes of spin transport in the Haldane-Shastry chain. We do this by considering time evolution from inhomogeneous perturbations of equilibrium states with inverse temperature $\beta$ and applied magnetic field $h$, i.e. inhomogeneous initial magnetic fields of the form
\begin{equation}
\label{eq:inhomIC}
h(x,0) = h + \delta h(x)
\end{equation}
where $\beta |\delta h (x)| \ll 1$ for all $x$. We let $\varsigma_{\mathrm{eq}}(k)$ and $s^z_{\mathrm{eq}}$ denote the equilibrium values of $\varsigma(k)$ and $s^z$ respectively. Then in the local density approximation for $\varsigma(x,k,0)$, it follows by combining Eqs. \eqref{eq:magicformula} and \eqref{eq:defmu} that the initial condition for the signed spinon density is given by
\begin{equation}
\varsigma(x,k,0) = \varsigma_{\mathrm{eq}}(k) + \frac{1}{2\pi}\frac{\tanh{\beta h \mu(k)}}{\tanh{\beta h}\cosh^2{\beta h \mu(k)}}\beta \delta h(x) + \mathcal{O}(\delta h^2).
\end{equation}
By Eqs. \eqref{eq:stimeevo} and \eqref{eq:magtimeevo}, the local magnetization density
\begin{equation}
s^z(x,t) = s^z_{\mathrm{eq}} + \int_{-\pi}^\pi \frac{dk}{4\pi} \frac{\tanh{\beta h \mu(k)}}{\tanh{\beta h}\cosh^2{\beta h \mu(k)}} \beta \delta h(x-Jkt/2) + \mathcal{O}(\delta h^2).
\end{equation}
and similarly the local spin current density
\begin{equation}
j^S(x,t) =  \int_{-\pi}^\pi \frac{dk}{4\pi} \frac{(Jk/2)\tanh{\beta h \mu(k)}}{\tanh{\beta h}\cosh^2{\beta h \mu(k)}} \beta \delta h(x-Jkt/2) + \mathcal{O}(\delta h^2).
\end{equation}
\subsection{Time evolution from weak domain walls}
\begin{figure}[t]
    \centering 
    \includegraphics[width=0.45\linewidth]{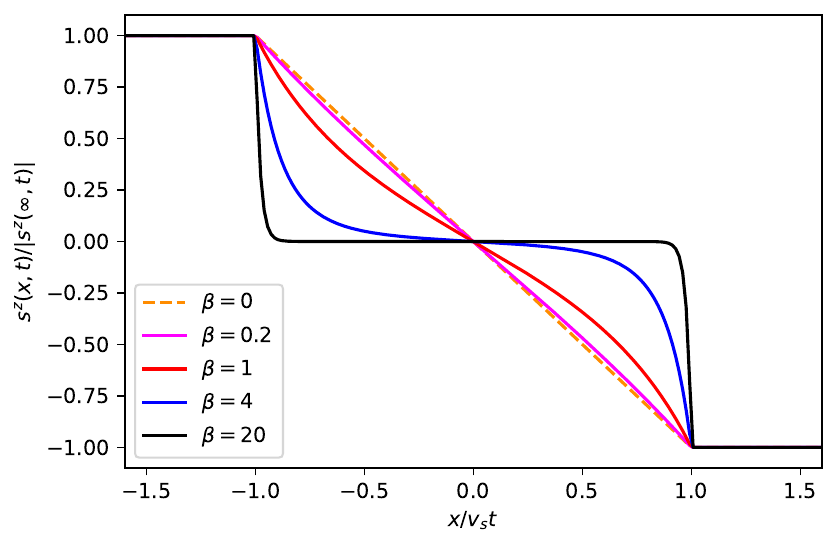}
        \includegraphics[width=0.45\linewidth]{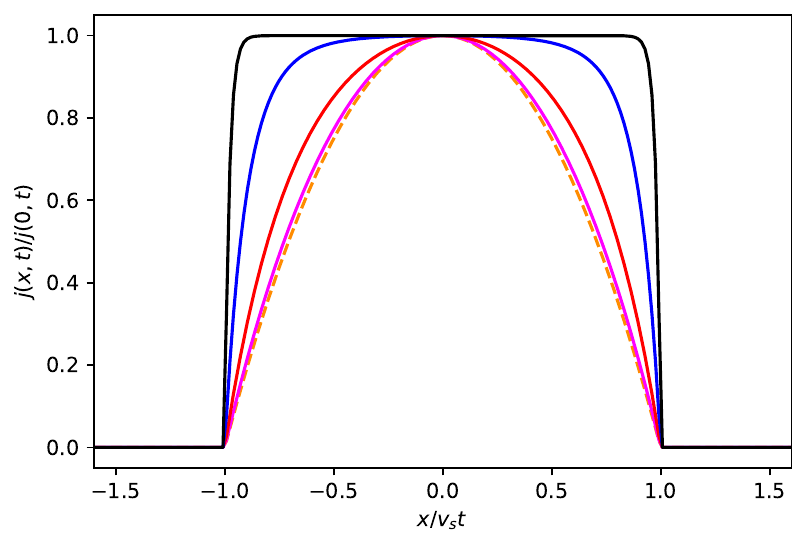}
    \caption{Hydrodynamic predictions for time evolution from ``weak domain wall'' initial conditions with a homogeneous bulk inverse temperature $\beta$ and with the left and right reservoirs prepared in equal and opposite magnetic fields $\beta h_L = -\beta h_R = \tilde{\epsilon} \ll 1$. The same legend pertains to both plots, and the crossover to the infinite temperature prediction Eq. \eqref{eq:tworesinft} as $\beta \to 0$ is clear.
    \textit{Left:} limiting $\tilde{\epsilon}\to 0$ spin density profiles predicted by Eq. \eqref{eq:tworeshalffill} as a function of $\beta$.
    \textit{Right:} limiting $\tilde{\epsilon}\to 0$ spin current density profiles predicted by Eq. \eqref{eq:tworeshalffill} as a function of $\beta$.}

    \label{Fig:twores}
\end{figure}

We first consider time evolution from the weak domain wall initial condition discussed in Sec. \ref{sec:SDW}. In Eq. \eqref{eq:inhomIC}, this can be written as the perturbation
\begin{equation}
\delta h(x) = \epsilon(1-2\Theta(x)),
\end{equation}
implying the inhomogeneous spin density and spin current density profiles
\begin{equation}
s^z(x,t) = -\tilde{\epsilon} \int_0^{\frac{\pi x}{v_s t}} \frac{dk}{2\pi} \frac{\tanh{\beta h \mu(k)}}{\tanh{\beta h}\cosh^2{\beta h \mu(k)}} + \mathcal{O}(\tilde{\epsilon}^2), \quad |x| \leq v_s t,
\end{equation}
and
\begin{equation}
\label{eq:tworescurr}
j(x,t) = -\tilde{\epsilon}  \int_{-\pi}^{\frac{\pi x}{v_st}}\frac{dk}{2\pi} \frac{(Jk/2)\tanh{\beta h \mu(k)}}{\tanh{\beta h}\cosh^2{\beta h \mu(k)}} + \mathcal{O}(\tilde{\epsilon}^2), \quad |x| \leq v_s t,
\end{equation}
within the quasiparticle lightcone (recall that $\tilde{\epsilon} = \beta \epsilon$). These expressions simplify in the half-filled and infinite temperature limits, as follows. At half filling $h=0$ we find that
\begin{equation}
\label{eq:tworeshalffill}
s^z(x,t) \sim -\frac{\tilde{\epsilon}}{2\sqrt{\beta v_s}} e^{-\frac{\beta \pi v_s}{4}}\mathrm{erfi}\left(\sqrt{\frac{\beta \pi v_s}{4}} \frac{x}{v_s t}\right), \quad j(x,t) \sim \frac{\tilde{\epsilon}}{\beta \pi} \left(1-e^{-\frac{\beta \pi v_s}{4}\left(1-\left(\frac{x}{v_st}\right)^2\right)}\right), \quad |x| \leq v_s t, \, \tilde{\epsilon} \to 0,
\end{equation}
where $\mathrm{erfi}(x) = \frac{2}{\sqrt{\pi}} \int_{0}^x dy \, e^{y^2}$ denotes the imaginary error function. Meanwhile in the limit $\beta \to 0$ with $\tilde{h} = \beta h = \mathrm{const}$, we find the following striking \textit{linear} profile for the steady-state magnetization:
\begin{align}
\label{eq:tworesinft}
s^z(x,t) \sim -\frac{\tilde{\epsilon}}{2 \cosh^2{\tilde{h}}} \frac{x}{v_st}, \quad j(x,t) \sim  \frac{\tilde{\epsilon}v_s}{4 \cosh^2{\tilde{h}}} \left(1-\left(\frac{x}{v_st}\right)^2\right) \quad |x| \leq v_s t, \, \tilde{\epsilon} \to 0.
\end{align}
The convergence in shape of the limiting $\tilde{\epsilon}\to 0$ profiles in Eq. \eqref{eq:tworeshalffill} to those in Eq. \eqref{eq:tworesinft} with increasing temperature is shown in Fig. \ref{Fig:twores}. There is a crossover from perfectly flat to perfectly linear magnetization profiles as the temperature increases. This prediction of perfectly linear magnetization profiles at infinite temperature is consistent with recent tensor-network-based numerical simulations of the Haldane-Shastry model~\cite{inprep}.

As usual, we expect that the non-differentiable features at the lightcone $|x| = v_s t$ will be smoothed out in practice by higher-order (presumably subdiffusive~\cite{Bulchandani_2019}) corrections to the kinetic equation Eq. \eqref{eq:BB}. In order to independently corroborate the analytical predictions Eqs. \eqref{eq:tworeshalffill} and \eqref{eq:tworesinft}, we now verify that they yield exact results for the spin Drude weight at zero and infinite temperature.

\subsection{Computing the spin Drude weight}
In principle, Eqs. \eqref{eq:tworescurr} and Eq. \eqref{eq:hydroDrude} can be used to efficiently compute the spin Drude weight at any temperature and any filling fraction. For concreteness, we focus on the two analytically tractable cases implied by Eq. \eqref{eq:tractlim}.
\subsubsection{Half filling}
Integrating the current in Eq. \eqref{eq:tworeshalffill}, we find that
\begin{equation}
\lim_{\tilde{\epsilon}\to 0} \frac{1}{\tilde{\epsilon}}\int_{-\infty}^{\infty} dx \, j(x,t) = \frac{2v_s t}{\beta \pi}\left(1 - \frac{1}{\sqrt{\beta v_s}}  e^{-\frac{\beta \pi v_s}{4}} \mathrm{erfi}\left(\sqrt{\frac{\beta \pi v_s}{4}}\right)\right).
\end{equation}
We deduce by Eq. \eqref{eq:hydroDrude} that
\begin{equation}
D^S(\beta,h=0) = \frac{v_s}{2 \beta \pi}\left(1 - \frac{1}{\sqrt{\beta v_s}}  e^{-\frac{\beta \pi v_s}{4}} \mathrm{erfi}\left(\sqrt{\frac{\beta \pi v_s}{4}}\right)\right).
\end{equation}
In the zero temperature limit $T \to 0$ this vanishes, which is consistent with the vanishing of the right-hand side of Eq. \eqref{eq:microDrude} in the ground state\cite{Bernevig01}. However, the rescaled Drude weight $\tilde{D}^S = \beta D^S$ does not vanish, and instead satisfies
\begin{equation}
\label{eq:lowTDrude}
\tilde{D}^S(\beta,h=0) \to \frac{v_s}{2\pi}, \quad T \to 0.
\end{equation}
This should be compared to the Luttinger liquid result, which states\cite{giamarchi2003quantum,Bertini2021RMP} for our choice of conventions that $\tilde{D}^S(\beta=\infty,h=0) = \frac{uK}{\pi}$, where $K$ denotes the Luttinger parameter and $u$ the quasiparticle velocity. Making the identifications~\cite{HaldaneOrig,ShastryOrig,stephan2017full} $u=v_s$ and $K=1/2$ for the Haldane-Shastry ground state recovers Eq. \eqref{eq:lowTDrude}.

\subsubsection{Infinite temperature}
Integrating the current in Eq. \eqref{eq:tworesinft}, we find that
\begin{equation}
\lim_{\tilde{\epsilon}\to 0} \frac{1}{\tilde{\epsilon}}\int_{-\infty}^{\infty} dx \, j(x,t) = \frac{v_s^2 t}{3 \cosh^2{\tilde{h}}}.
\end{equation}
It follows by Eq. \eqref{eq:hydroDrude} that
\begin{equation}
\label{eq:hydroDrudeinfT}
D^S(\beta,h) = \frac{v_s^2}{12 \cosh^2{\tilde{h}}}, \quad \tilde{h} = \beta h, \, \beta \to 0.
\end{equation}
However, at infinite temperature we can evaluate the thermal expectation value of the square of the rapidity operator in Eq. \eqref{eq:microDrude} exactly, to yield
\begin{equation}
\langle \hat{\Lambda}^2 \rangle_0 =  \sum_{x \neq y} \left(\frac{\pi}{L}\right)^2 \cot^2{\left(\frac{\pi(x-y)}{L}\right)}\left(\frac{1}{4} - \langle \hat{S}^z_x \hat{S}^z_y \rangle_0 \right), \quad \tilde{h} = \beta h, \, \beta \to 0,
\end{equation}
so that an exact expression for the infinite-temperature spin Drude weight of the Haldane-Shastry chain is given by
\begin{equation}
\label{eq:infTexact}
D^S(\beta,h) = \frac{J^2}{2} \sum_{n=1}^\infty \frac{1}{n^2}\left(\frac{1}{4}-\langle \hat{S}_0^z \hat{S}_n^z\rangle_0\right) = \frac{J^2}{8} \frac{\zeta(2)}{\cosh^2{\tilde{h}}} = \frac{v_s^2}{12 \cosh^2{\tilde{h}}}, \quad \tilde{h} = \beta h, \, \beta \to 0,
\end{equation}
which confirms the hydrodynamic prediction Eq. \eqref{eq:hydroDrudeinfT}.
\subsection{Hydrodynamics of the spin structure factor}

If we now let $\delta h(x) = \frac{\epsilon}{2}e^{iqx}$, corresponding to the choice $f(x) = e^{iqx}$ in Section \ref{sec:DSF}, we obtain the prediction
\begin{equation}
K(q,t) = \frac{1}{\beta \epsilon} e^{-iqx} s^z(x,t) = \int_{-\pi}^\pi \frac{dk}{8\pi} \frac{\tanh{\beta h \mu(k)}}{\tanh{\beta h}\cosh^2{\beta h \mu(k)}} e^{-iqJkt/2}, \quad |q| \ll 1,
\end{equation}
for the long-wavelength dynamics of the Kubo correlation function. Fourier transforming in time yields
\begin{equation}
K(q,\omega) = \frac{1}{2 |q| J} \Theta\left(v_s^2 q^2-\omega^2\right) \frac{\tanh{\beta h \mu(\pi\omega/v_s q)}}{\tanh{\beta h}\cosh^2{\beta h \mu(\pi \omega / v_s q)}}, \quad |q| \ll 1,
\end{equation}
where $\Theta(x)$ denotes the Heaviside step function. We deduce by Eq. \eqref{eq:FourierRel} that the spin structure factor
\begin{equation}
\label{eq:SSF}
C(q,\omega) = \frac{1}{2 |q|J}\left(\frac{\beta \omega}{1-e^{-\beta \omega}}\right) \Theta\left(v_s^2 q^2-\omega^2\right)  \frac{\tanh{\beta h \mu(\pi\omega/v_s q)}}{\tanh{\beta h}\cosh^2{\beta h \mu(\pi \omega / v_s q)}}, \quad |q| \ll 1,
\end{equation}
in Fourier space. This expression can be simplified in various limits that we now discuss.

\subsubsection{Half filling}
Let us first consider the case of half filling. Then Eq. \eqref{eq:tractlim} implies that
\begin{equation}
\label{eq:halffilledC}
C(q,\omega) = \frac{1}{2 |q|J}\left(\frac{\beta \omega}{1-e^{-\beta \omega}}\right) \Theta\left(v_s^2 q^2-\omega^2\right)\exp{\left(-\frac{\beta}{2J} \left(v_s^2-\frac{\omega^2}{q^2}\right)\right)}, \quad |q| \ll 1.
\end{equation}
We note that the dynamical spin structure factor at half filling and zero temperature was previously computed by Haldane and Zirnbauer using a mapping to random matrix theory\cite{HaldaneZirnbauer,Zaletel_2015}. Their expression for the spin structure factor is dominated by its oscillatory prefactor at $q=\pi$, which is presumably beyond hydrodynamics, but is also supported~\cite{HaldaneTalstra} in the vicinity of $q=0$, which should be within the scope of hydrodynamics.

In Eq. \eqref{eq:halffilledC}, the strictly zero temperature limit corresponds to taking $\beta \omega \gg 1$, also known as the collisionless or phase-coherent regime of quantum transport\cite{Damle_1997}. In order to compare to the Haldane-Zirnbauer formula, we first work in the collisionless regime before taking $T \to 0$, which yields
\begin{equation}
C(q,\omega) \sim \frac{1}{|q|}  \Theta(v_s^2 q^2 -\omega^2) \left(\frac{\beta \omega}{2J}\right) \exp{\left(-\frac{\beta\omega}{2J} \left(\frac{v_s^2}{\omega}-\frac{\omega}{q^2}\right)\right)}, \quad |q| \ll 1.
\end{equation}
Taking $T \to 0$ in this expression, the exponential gives rise to a (one-sided) delta function,
\begin{equation}
\label{eq:hydrodelta}
C(q,\omega) = \frac{1}{2|q|} \delta\left(\frac{v_s^2}{\omega} - \frac{\omega}{q^2}\right)
=\frac{|q|}{4}\left(\delta(\omega-v_s q) + \delta(\omega+v_s q)\right)
, \quad |q| \ll 1, T=0.
\end{equation}

We now compare this expression with the ``hydrodynamic limit'' of the Haldane-Zirnbauer\cite{HaldaneZirnbauer} formula for the dynamical spin structure factor of the Haldane-Shastry chain at $T=0$ and half filling. In the limit of zero temperature and with our Fourier conventions, the Haldane-Zirnbauer formula reads
\begin{equation}
    C_{\mathrm{HZ}}(q,\omega) = \frac{\pi}{4} \frac{\Theta(E_2(Q)-\omega) \Theta(\omega-E_{1-}(Q))\Theta(\omega-E_{1+}(Q))}{\sqrt{(\omega-E_{1-}(Q))(\omega-E_{1+}(Q))}}
     \label{eq:HaldaneZirnbauer}
\end{equation}
where
\begin{equation}
    E_2(Q) = \frac{v_s}{2\pi}Q(2\pi-Q),
    \quad
    E_{1-}(Q) =\frac{v_s}{\pi}Q(\pi-Q),
    \quad
    E_{1+}(Q)=\frac{v_s}{\pi}(Q-\pi)(2\pi-Q).
\end{equation}

The range of crystal momenta that Haldane and Zirnbauer consider is $Q \in [0,2\pi)$, which is related to our choice of crystal momenta $q \in [-\pi,\pi)$ by the change of variables $Q(q) = q$ for $0 \leq Q < \pi$ and $Q(q) = Q-2\pi$ for $\pi \leq q < 2 \pi$. Since the Haldane-Zirnbauer formula is even in $q$, we will focus on $q>0$ without loss of generality. In the long-wavelength limit $ 0 < q \ll 1$, we have $E_2(q), E_{1-}(q) \to 0$ and $E_{1+}(q) \to - 2\pi v_s$. Thus for small $|q|$, Eq. \eqref{eq:HaldaneZirnbauer} is dominated by the singular contributions due to $E_2(q)$ and $E_{1-}(q)$ and we have
\begin{equation}
    C_{\mathrm{HZ}}(q,\omega) \approx \frac{1}{4} \sqrt{\frac{\pi}{2v_s}} \frac{\Theta(E_2(q)-\omega)\Theta(\omega-E_{1-}(q))}{\sqrt{\omega-E_{1-}(q))}}.
    \label{simplify_HaldaneZirnbauer}
\end{equation}
Notice that Eq.~\eqref{simplify_HaldaneZirnbauer} only supported in the region $E_{1-}(q) \leq \omega \leq E_2(q)$, which is concentrated about the spinon dispersion $\omega=v_s q$ with a width $v_sq^2/2\pi \to 0$ that vanishes as $q \to 0$. Moreover, for a fixed value of $q$,
\begin{equation}
    \int_{-\infty}^{\infty} d\omega \, C_{\mathrm{HZ}}(q,\omega) 
    = \int_{E_{1-}(q)}^{E_2(q)} d\omega \, C_{\mathrm{HZ}}(q,\omega)
     \approx \frac{1}{4} \sqrt{\frac{\pi}{2v_s}} \int_{v_s q - (v_s/\pi)q^2}^{v_s q - (v_s/2\pi)q^2} d \omega \frac{1}{\sqrt{\omega-v_s q + (v_s/\pi)q^2}}
    =\frac{q}{4}.
\end{equation}
Thus the Haldane-Zirnbauer formula behaves like a nascent delta function in $\omega$ with weight $q/4$ at long wavelengths, and we deduce that 
\begin{equation}
C_{\mathrm{HZ}}(q,\omega) \approx \frac{q}{4} \delta(\omega-v_sq), \quad  0 < q \ll 1,
\end{equation}
in agreement with the hydrodynamic prediction Eq. \eqref{eq:hydrodelta}.

\subsubsection{Infinite temperature}
\begin{figure}[t]
    \centering 
    \includegraphics[width=0.45\linewidth]{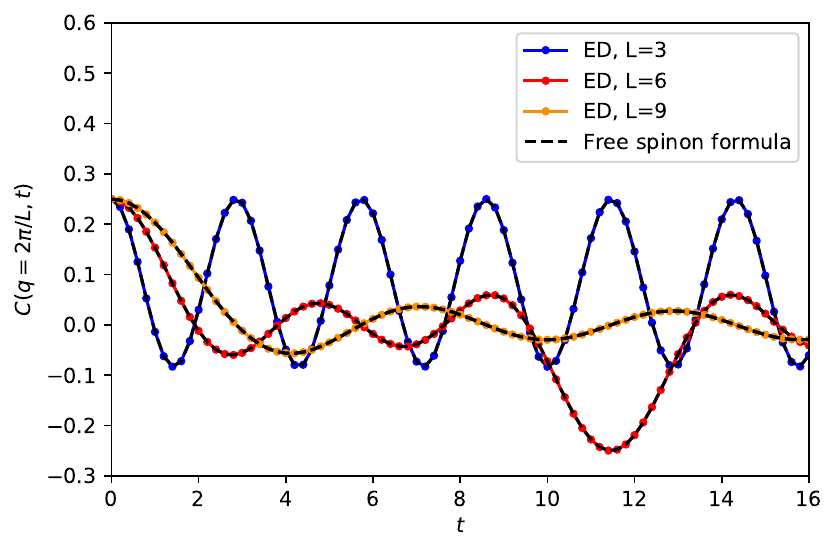}
        \includegraphics[width=0.45\linewidth]{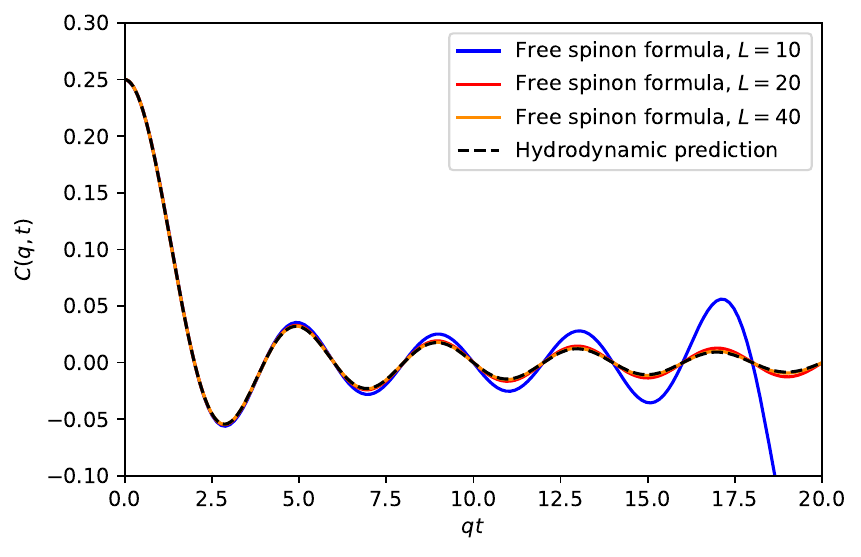}
    \caption{\textit{Left:} dynamics of the longitudinal spin structure factor at infinite temperature, zero magnetization and the longest possible wavelength $q=2\pi/L$ obtained from exact diagonalization for $L=3,6,9$, versus the ``free spinon'' formula Eq. \eqref{eq:emp}. \textit{Right:} convergence of the free spinon formula Eq. \eqref{eq:emp} to the hydrodynamic prediction Eq. \eqref{eq:BAinfT} with increasing system size.}

    \label{fig:empirical}
\end{figure}
We next consider the limit of infinite temperature $\beta \to 0$ with the filling held fixed, $\beta h = \tilde{h} = \mathrm{const}$. In Fourier space, this yields
\begin{equation}
C(q,\omega) \sim \frac{1}{\cosh^2{\tilde{h}}}\frac{1}{2|q|J} \Theta\left(v_s^2 q^2-\omega^2\right)  , \quad |q| \ll 1,
\end{equation}
by Eq. \eqref{eq:defmu}. To obtain real time and space dynamics, it is simplest to note that $C(q,t)=K(q,t)$ at infinite temperature by Eq. \eqref{eq:FourierRel}, i.e. that the Kubo correlation function coincides with the dynamical structure factor\cite{Bertini2021RMP}. Thus we obtain the simple, explicit expression
\begin{equation}
\label{eq:BAinfT}
C(q,t)  =\frac{1}{\cosh^2{\tilde{h}}} \int_{-\pi}^\pi \frac{dk}{8\pi} e^{-iqJkt/2} = \frac{1}{4 \cosh^2{\tilde{h}}} \frac{\sin{\left(v_s q t \right)}}{v_s q t}
, \quad q \ll 1.
\end{equation}
Fourier inverting this expression in space predicts a ballistically expanding ``boxcar'' profile in real space, namely
\begin{equation}
C(x,t) \approx \frac{1}{4 \cosh^2{\tilde{h}}} \frac{1}{2v_s t}\left(\Theta(x+v_s t) - \Theta(x-v_s t)\right), \quad x,t \gg 1.
\end{equation}
As a nontrivial test of Eq. \eqref{eq:BAinfT}, we have found that at infinite temperature and zero magnetization, $\tilde{h}=0$, the following ``free spinon'' formula 
\begin{equation}
\label{eq:emp}
C_{\mathrm{emp.}}(q = 2\pi/L,t) = \frac{1}{4L} \sum_{k} \cos{\left((e_0(k) - e_0(k+q))t\right)}
\end{equation} 
yields the dynamical spin structure factor of the finite Haldane-Shastry chain at the longest possible wavelength. We have found empirically that this formula is exact for all system sizes and times, see left panel of Fig. \ref{fig:empirical} for a comparison with exact diagonalization.
In the thermodynamic limit, we have
\begin{equation}
C_{\mathrm{emp.}}(q = 2\pi/L,t) \sim \frac{1}{4}\frac{\sin{\left(v_s q t \right)}\cos{\left(Jq^2t/4\right)}}{v_s q t}  = \frac{1}{4}\frac{\sin{\left(v_s q t \right)}}{v_s q t} + \mathcal{O}(q^4), \quad L \to \infty,
\end{equation}
which matches the hydrodynamic prediction Eq. \eqref{eq:BAinfT} at leading non-trivial order in $q$.  Convergence of the empirically exact result Eq. \eqref{eq:emp} to the hydrodynamic prediction Eq. \eqref{eq:BAinfT} as a function of $qt$ is shown in the right panel of Fig. \ref{fig:empirical}, and serves as a nontrivial microscopic test of the hydrodynamic result. We defer theoretical justification for Eq. \eqref{eq:emp} to future work. 

\section{Conclusion}
We have shown that a kinetic theory of Bethe quasiparticles yields a convincing account of the hydrodynamics of the Haldane-Shastry chain. Several further questions now arise.

The first is the microscopic derivation of the system of Boltzmann equations Eq. \eqref{eq:BB}. At this point, there exist several possible microscopic routes towards deriving kinetic equations for integrable systems\cite{borsi2021current,cubero2021form,Bulchandani_2021,Bertini_2022,doyon2023abinitioderivationgeneralised,spohn2024hydrodynamic}. A particularly intuitive and direct derivation is possible for integrable systems that possess an extensively large Lax matrix. In such cases, it is possible to formulate a ``Bethe-Lax correspondence''~\cite{BulchandaniToda,Bulchandani_2021} between the spectrum of the Lax matrix and the quasiparticle dynamics, an idea that has been successfully applied to the Toda lattice~\cite{BulchandaniToda,DoyonToda,spohn2021hydrodynamicequationstodalattice}, to Calogero models~\cite{Bulchandani_2021,spohn2024hydrodynamic} and to the Ablowitz-Ladik chain~\cite{Spohn_2022,brollo2024particlescatteringfusionablowitzladik,spohn2024currentsablowitzladikchain}. While most of these derivations pertain to classical systems, they can be extended to the hydrodynamics of the quantum Toda lattice~\cite{BulchandaniToda,spohn2021hydrodynamicequationstodalattice} and we have similarly checked that the quantum Lax matrix can be used to obtain the kinetic theory of the quantum Calogero model~\cite{Bulchandani_2021}. Given the intimate connection between Calogero models and the Haldane-Shastry chain~\cite{PolychronakosReview}, it is tempting to expect that such approaches should extend to the Haldane-Shastry chain as well. However, the structure of conservation laws for the Haldane-Shastry chain is somewhat unconventional~\cite{inozemtsev1990connection,Yangian,haldane1994physics,TalstraConsLaws} compared to other integrable models, and we leave the existence of a Bethe-Lax correspondence for the Haldane-Shastry chain as an open question.

Another striking feature of the equations Eq. \eqref{eq:BB} compared to more conventional integrable systems is the lack of velocity dressing. This has its origin in the possibility of interpreting this model as a non-interacting spinon gas~\cite{haldane1991spinon,EsslerSpinon,NoSpinonAttraction}. In agreement with expectations for non-interacting particles, Eq. \eqref{eq:BB} predicts ballistic transport for all the conserved charges of the Haldane-Shastry model, including spin, energy and all higher conserved charges associated with integrability~\cite{TalstraConsLaws}. However, a full explanation of this pervasive ballistic transport seems to require an exact microscopic mapping from the Haldane-Shastry chain to a gas of non-interacting spinons, whose construction has remained an open question for some thirty years~\cite{TalstraSpinon}. A more modest goal would be to understand such ballistic transport at the operatorial level; Eq. \eqref{eq:BB} superficially implies that every local conserved charge of the Haldane-Shastry chain gives rise to a conserved current operator, which demands an algebraic explanation if this is indeed the case.

Finally, it is worth noting that the Haldane-Shastry chain is an extremal point of the family of $SU(2)$-symmetric spin-$1/2$ chains discovered by Inozemtsev~\cite{inozemtsev1990connection}, that interpolate between the inverse-square-interacting Haldane-Shastry model and the nearest-neighbour-interacting Heisenberg chain. The fact that spin transport in the Haldane-Shastry chain is purely ballistic, while spin transport in the Heisenberg chain is not, raises two puzzling questions. First, how does spin transport vary across the family of Inozemtsev models? Presumably spin transport becomes anomalous as soon as the spin-spin interaction length becomes finite, as has been found for other $SU(2)$-symmetric integrable spin chains~\cite{ZnidOrig,KPZ,dupont2020universal,Ilievski_2021}. If this intuition is correct, the question then arises of how the nature of spin transport reflects the structure of the underlying quantum group. The non-trivial $S$-matrix for quasiparticle scattering in the hyperbolic Inozemtsev models~\cite{klabbers2016thermodynamics} suggests that spinons in these models interact with one another, unlike for the Haldane-Shastry chain~\cite{EsslerSpinon}, and that such interactions play a role in the emergence of anomalous spin transport.

\section{Acknowledgments}
We thank S. Gopalakrishnan, D.A. Huse and J.E. Moore for helpful discussions, S. Anand and J. Kemp for sharing their numerical results before publication and F.D.M. Haldane for suggesting this problem.
\bibliography{bibl.bib}
\end{document}